\def\year{2020}\relax
\documentclass[letterpaper]{article} %
\usepackage{aaai20}  %
\usepackage{times}  %
\usepackage{helvet} %
\usepackage{courier}  %
\usepackage[hyphens]{url}  %
\usepackage{graphicx} %
\usepackage{amsmath}
\usepackage{amssymb}
\usepackage{subfigure}
\usepackage{braket}
\usepackage{enumitem}

\newcommand{\mat}[1]{{\bf #1}}   %
\newcommand{\ourmethod}{\textsc{AHN}}
\newcommand{\citet}[1]{\citeauthor{#1} \shortcite{#1}}

\title{Asymmetrical Hierarchical Networks with Attentive Interactions for Interpretable Review-Based Recommendation}

\author{Xin Dong,\textsuperscript{\rm 1}\thanks{This work was done when the first author was an intern at NEC Laboratories America. {$^\ddagger$}Corresponding author.}
Jingchao Ni,\textsuperscript{\rm 2}{$^\ddagger$}
Wei Cheng,\textsuperscript{\rm 2}
Zhengzhang Chen,\textsuperscript{\rm 2}
Bo Zong,\textsuperscript{\rm 2}\\
\bf\Large{Dongjin Song,\textsuperscript{\rm 2}
Yanchi Liu,\textsuperscript{\rm 2}
Haifeng Chen,\textsuperscript{\rm 2}
and Gerard de Melo\textsuperscript{\rm 1}}\\
\textsuperscript{\rm 1}Rutgers University,
\textsuperscript{\rm 2}NEC Laboratories America\\
\textsuperscript{\rm 1}\{xd48@rutgers.edu, gdm@demelo.org\}\\
\textsuperscript{\rm 2}\{jni, weicheng, zchen, bzong, dsong, yanchi, haifeng\}@nec-labs.com}

\begin{document}

\maketitle

\begin{abstract}
Recently, recommender systems have been able to emit substantially improved recommendations by leveraging user-provided reviews. Existing methods typically merge all reviews of a given user or item into a long document, and then process user and item documents in the same manner. In practice, however, these two sets of reviews are notably different: users' reviews reflect a variety of items that they have bought and are hence very heterogeneous in their topics, while an item's reviews pertain only to that single item and are thus topically homogeneous. In this work, we develop a novel neural network model that properly accounts for this important difference by means of asymmetric attentive modules. The user module learns to attend to only those signals that are relevant with respect to the target item, whereas the item module learns to extract the most salient contents with regard to properties of the item. Our multi-hierarchical paradigm accounts for the fact that neither are all reviews equally useful, nor are all sentences within each review equally pertinent. Extensive experimental results on a variety of real datasets demonstrate the effectiveness of our method.
\end{abstract}

\section{Introduction}
The rapid shift from traditional retail and services to online transactions has brought forth a large volume of review data in areas such as e-commerce, dining, tourism, among many others. While such reviews are routinely consulted directly by consumers and affect their decision making, recent work has shown that they can also be exploited by intelligent algorithms. The detailed semantic cues that they harbor not only reveal different aspects (\textit{e.g.}, quality, material, color, \textit{etc.}) of an item, but also reflect the sentiment of users towards these aspects. Such fine-grained signals are extremely valuable to a recommender system and significantly complement the sparse rating and click-through data, based on which many traditional collaborative filtering methods \cite{koren2009matrix} have been developed. Thus, there has been a series of studies seeking to harness the potential of reviews in improving the recommendation quality \cite{zheng2017joint,catherine2017transnets,seo2017interpretable,chen2018neural}.

\begin{figure}[!t]
\includegraphics[width=1.0\columnwidth]{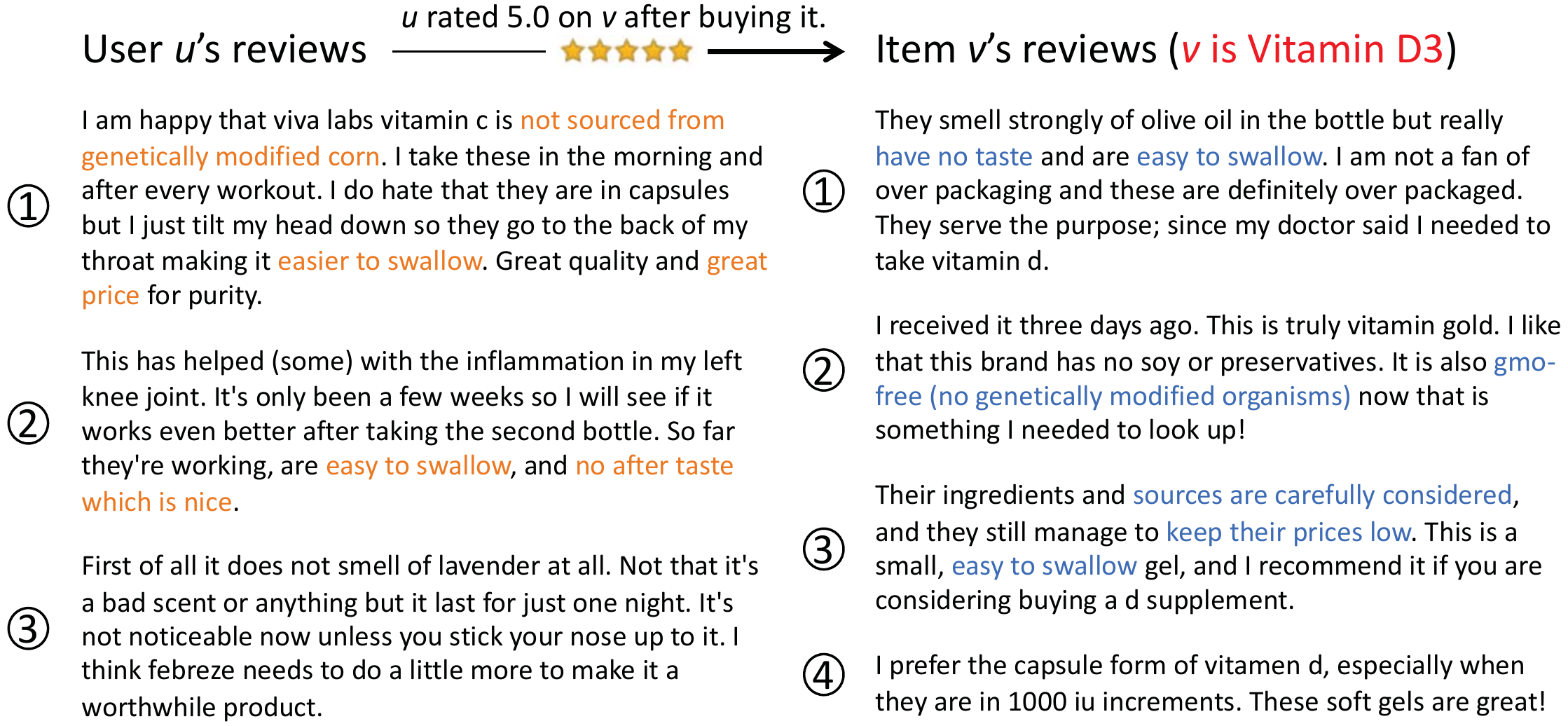}
\caption{An example of reviews by a user and for an item. User $u$ rated a 5.0 score on item $v$ after purchasing it.}\label{fig.intro}
\end{figure}

These studies have shown that leveraging reviews can indeed boost the recommendation effectiveness quite remarkably. Typically, they associate users with the respective sets of reviews they have written, while associating each item with the set of all reviews that have been written for it. To predict the rating for an unseen user--item pair, in a first step, the embeddings of that user and item are inferred from the respective sets of reviews via a neural network. Then, the two embeddings are matched to predict a numeric rating between them. For example, DeepCoNN \cite{zheng2017joint} relies on convolutional neural networks to learn user (item) embeddings, and on a factorization machine \cite{rendle2010factorization} to predict ratings. D-ATT \cite{seo2017interpretable} uses dual-attention based networks to learn embeddings, and a simple dot product to predict ratings.

Despite the encouraging progress, existing methods all regard the set of reviews by a user and the set of reviews for an item as the same type of documents, and invoke the same model (or even a shared model) to process them in parallel. %
In reality, however, the set of reviews for a user is fundamentally different from the set of reviews for an item. In particular, reviews for users correspond to a diverse set of items that they have rated, resulting in notably {\em heterogeneous} textual contents with a variety of topics for different items. In contrast, each item's reviews are only about itself, and the contents are thus {\em homogeneous} in the sense that the topic is limited to a single narrow domain. For example, Fig.~\ref{fig.intro} shows several reviews from Amazon's health domain. User $u$'s historical reviews describe three items, Vitamin C, anti-inflammatory medication, and an air freshener, while all reviews for item $v$ are about itself, \textit{i.e.}, Vitamin D3.

This profound difference necessitates distinct forms of attention to be paid on user reviews as opposed to item reviews, when deciding whether to recommend an item $v$ to a user $u$. To predict $u$'s preference of $v$, it is important to extract from $u$'s reviews those aspects that pertain most to $v$, e.g., comments on items that are similar to $v$. In contrast, from $v$'s reviews, we wish to account for the sentiment of other users with regard to  relevant aspects of $v$. If $u$ pays special attention to certain aspects of items similar to $v$, while other users wrote highly about $v$ with regard to these particular aspects, then it is much more likely that $v$ will be of interest to $u$. For example, in Fig.~\ref{fig.intro}, reviews 1 and 2 of $u$ are about non-prescription medicines that are similar to $v$. In reviews 1 and 2, $u$ mentioned aspects such as ``not sourced from genetically modified corn'', ``easier to swallow'', ``great price'', and ``no after taste'', indicating that $u$ considers the source and price and prefers easily swallowed products without after-taste. Meanwhile, reviews 1-3 of $v$ mention that $v$ ``have no taste'', is ``easy to swallow'', ``gmo-free'', and ``prices low'', which are opinions expressed by others that match $u$'s preferences. Thus, $v$ is likely to be of interest to $u$, and $u$ indeed marked a 5.0 score on $v$ after purchasing it.

Another vital challenge is how to reliably represent each review. Importantly, sentences are not equally useful within each review. For example, in Fig.~\ref{fig.intro}, the second sentence in $u$'s review 1, ``\emph{I take these in the morning and after every workout.}'' conveys little regarding $u$'s concerns for Vitamin C, and thus is less pertinent than other sentences in the same review. Since including irrelevant sentences can introduce noise and may harm the final embedding quality, it is crucial to aggregate only useful sentences to represent each review.

To address the above challenges, in this paper, we propose an \underline{A}symmetrical \underline{H}ierarchical \underline{N}etwork with Attentive Interactions (\ourmethod) for recommendation. \ourmethod\ progressively aggregates salient sentences to induce review representations, and aggregates pertinent reviews to induce user and item representations. \ourmethod\ is particularly characterized by its asymmetric attentive modules to flexibly distinguish the learning of user embeddings as opposed to item embeddings. For items, several attention layers are invoked to highlight sentences and reviews that contain rich aspect and sentiment information. For users, we designed an interaction-based co-attentive mechanism to dynamically select a homogeneous subset of contents related to the current target item. In this manner, \ourmethod\ hierarchically induces embeddings for user--item pairs reflecting the most useful knowledge for personalized recommendation. In summary, our contributions are
\begin{enumerate}[noitemsep]
\item We identify the asymmetric attention problem for review-based recommendation, which is important but neglected by existing approaches.
\item We propose \ourmethod, a novel deep learning architecture that not only captures both of the asymmetric and hierarchical characteristics of the review data, while also enabling interpretability of the results.
\item We conduct experiments on 10 real datasets. The results demonstrate that \ourmethod\ consistently outperforms the state-of-the-art methods by a large margin, while providing good interpretations of the predictions.
\end{enumerate}

\section{Related Work}

Exploiting reviews 
has proven considerably useful in recent work on recommendation. %
Many methods primarily focus on topic modeling based on the review texts. For example, HFT \cite{mcauley2013hidden} employs LDA to discover the latent aspects of users and items from reviews. RMR \cite{ling2014ratings} extracts topics from reviews to enhance the user and item embeddings obtained by factorizing the rating matrix. TopicMF \cite{bao2014topicmf} jointly factorizes a rating matrix and bag-of-words representations of reviews to infer user and item embeddings. Despite the improvements achieved, these methods only focus on topical cues in reviews, but neglect the rich semantic contents. Moreover, they typically represent reviews as bag-of-words, and thus remain oblivious of the order and contexts of words and sentences in reviews, which are essential for modeling the characteristics of users and items \cite{zheng2017joint}.

Inspired by the astonishing advances of recent deep NLP techniques in various applications \cite{santos2016attentive,wang2018co,peters2018deep,dong2018helping,devlin2018bert,yang2019xlnet}, there has been increasing interest in studying deep learning models. DeepCoNN \cite{zheng2017joint} employs CNNs as an automatic feature extractor to encode each user and item into a low-dimensional vector by assessing the relevant set of historical reviews. TransNet \cite{catherine2017transnets} extends DeepCoNN by augmenting the CNN architecture with a multi-task learning scheme to regularize the user and item embeddings towards the target review. These methods, however, lack interpretability \cite{xian2019KGRecommendations} in their results.

To better understand the predictions, several attention-based methods have been developed. D-ATT \cite{seo2017interpretable} incorporates two kinds of attention mechanisms on the words of reviews to find informative words. NARRE \cite{chen2018neural} invokes review-level attention weights to aggregate review embeddings to form user (item) embeddings. HUITA \cite{wu2019hierarchical} is equipped with a symmetric hierarchical structure, where, at each level (e.g., word level), a regular attention mechanism is employed to infer the representation of the subsequent level (e.g., sentence level). MPCN \cite{tay2018multi} models the interactions between a user's reviews and an item's reviews via co-attention based pointers that are learned with the Gumbel-Softmax trick \cite{jang2016categorical}. However, all these methods just learn user and item embeddings in parallel and fail to consider the important differences between the two. As discussed before, this leads to suboptimal predictions.

Unlike the aforementioned methods, our method learns several hierarchical aggregators to infer user (item) embeddings. The aggregators are asymmetric to flexibly pay varying levels of attention to a user's (item's) reviews, so as to enhance the prediction accuracy and model interpretability.

\section{Our Proposed Model}

In this section, we introduce our \ourmethod\ model in a bottom-up manner. Fig.~\ref{fig.model} illustrates the architecture of \ourmethod.

\subsection{Sentence Encoding}

\begin{figure}[!t]
\includegraphics[width=1.0\columnwidth]{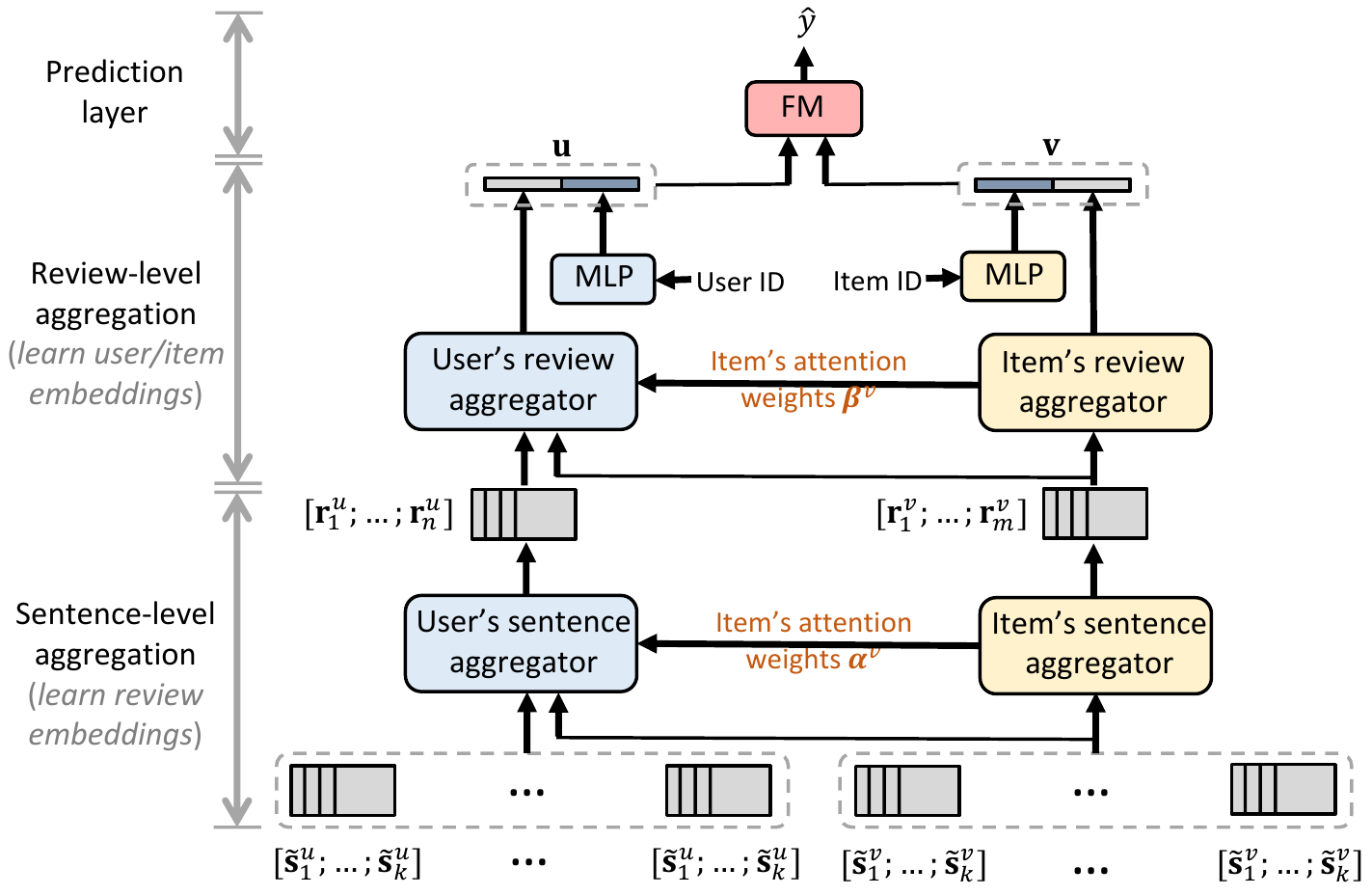}
\caption{The overall architecture of \ourmethod.}\label{fig.model}
\end{figure}

The sentence encoding layer (omitted in Fig.~\ref{fig.model}) aims to transform each sentence (in each review) from a sequence of discrete word tokens to a continuous vector embedding. We use a word embedding model to lay the foundation of this layer. Suppose the sentence $s$ has $l$ words. By employing a word embedding matrix $\mathbf{E} \in \mathbb{R}^{d \times |\mathcal{V}|}$, $s$ can be represented by a sequence $[\mathbf{e}_{1}, ..., \mathbf{e}_{l}]$, where $\mathbf{e}_{i}$ is the embedding of the $i$-th word in $s$, $d$ is the dimensionality of the word embedding, and $\mathcal{V}$ is the whole vocabulary of words. The matrix $\mathbf{E}$ can be initialized using word embeddings such as word2vec \cite{mikolov2013distributed} and GloVe \cite{pennington2014glove}, which are widely used in NLP. To refine the word embeddings, $\mathbf{E}$ is fine-tuned during model training.

To learn an embedding for $s$, we employ a bi-directional LSTM \cite{peters2018deep} on its constituent word embeddings, and apply max-pooling on the hidden states to preserve the most informative information. That is
\begin{equation}\label{eq.s}
    \begin{aligned}
        \mat{s} = \max([\mathbf{\tilde{e}}_{1}, ..., \mathbf{\tilde{e}}_{l}]),\\
    \end{aligned}
\end{equation}
where $\mat{s}$ is the embedding of $s$ and
\begin{equation}
    \begin{aligned}
        \mat{\tilde{e}}_{i} = \text{BiLSTM}(\mat{\tilde{e}}_{i-1}, \mat{e}_{i})~~(1 \le i \le l),
    \end{aligned}
\end{equation}
where $\mat{\tilde{e}}_{0}$ is initialized by an all-zero vector $\mat{0}$.

Suppose a review has $k$ sentences. We can then represent this review by a sequence $[\mat{s}_{1}, ..., \mat{s}_{k}]$, where $\mat{s}_{i}$ is the embedding of the $i$-th sentence in the review, as inferred by Eq.~\eqref{eq.s}. However, using Eq.~\eqref{eq.s}, each $\mat{s}_{i}$ only encodes its own semantic meaning, but remains oblivious of any contextual cues from its surrounding sentences in the same review. To further refine the sentence embedding, we introduce a context-encoding layer by employing another bi-directional LSTM on top of the previous layer to model the temporal interactions between sentences, i.e.,
\begin{equation}
    \begin{aligned}
        \mat{\tilde{s}}_{i} = \text{BiLSTM}(\mat{\tilde{s}}_{i-1}, \mat{s}_{i})~~(1 \le i \le k),
    \end{aligned}
\end{equation}
where $\mat{\tilde{s}}_{i}$ is the final embedding of the $i$-th sentence in the review and $\mat{\tilde{s}}_{0}$ is initialized as $\mat{0}$.

\subsection{Sentence-Level Aggregation}

Next, we develop sentence-level aggregators to embed each review into a compact vector from its constituent sentences. As discussed before, an ideal method should learn review embeddings in an asymmetric style. Thus, we design \ourmethod\ to learn different attentive aggregators for users and items, respectively, as highlighted in Fig.~\ref{fig.model}.

\subsubsection{Sentence Aggregator for Items.}
Given an item, we are interested in sentences that contain other users' sentiments on different aspects of the item, which are the key factors to determine its overall rating. To build an informative embedding for each review upon such sentences, we use a sentence-level attention network to aggregate the sentence embeddings $[\mat{\tilde{s}}_{1}^{v}, ..., \mat{\tilde{s}}_{k}^{v}]$ as follows, where the superscript $v$ is used to distinguish an item's notation from a user's notation.
\begin{equation}\label{eq.r.v}
    \begin{aligned}
        \mat{r}^{v} = \sum_{i=1}^{k}\alpha_{i}^{v}\mat{\tilde{s}}_{i}^{v},
    \end{aligned}
\end{equation}
Here, $\sum_{i=1}^{k}\alpha_{i}^{v}=1$, and $\alpha_{i}^{v}$ is the attention weight assigned to sentence $\mat{\tilde{s}}_{i}^{v}$. It quantifies the informativeness of sentence $\mat{\tilde{s}}_{i}^{v}$ with respect to $v$'s overall rating, compared to other sentences. The weights $\alpha_{i}^{v}$ are computed by our attentive module, taking the sentence embeddings as the input, as
\begin{equation}\label{eq.v.s.alpha}
    \begin{aligned}
        \alpha_{i}^{v} = \frac{\exp{(\mat{v}_{s}^{\top}(\tanh{(\mat{W}_{s}\mat{\tilde{s}}_{i}^{v})}\otimes\sigma(\mat{\hat{W}}_{s}\mat{\tilde{s}}_{i}^{v})))}}{\sum_{j=1}^{k}\exp{(\mat{v}_{s}^{\top}(\tanh{(\mat{W}_{s}\mat{\tilde{s}}_{j}^{v})}\otimes\sigma(\mat{\hat{W}}_{s}\mat{\tilde{s}}_{i}^{v})))}}.
    \end{aligned}
\end{equation}
Here, $\mat{v}_{s} \in \mathbb{R}^{h \times 1}$, $\mat{W}_{s} \in \mathbb{R}^{h \times d}$, and $\mat{\hat{W}}_{s} \in \mathbb{R}^{h \times d}$ are parameters, $\otimes$ is the element-wise product, and $\sigma(\cdot)$ is the sigmoid function. %
As suggested by \citet{ilse2018attention}, the approximate linearity of $\tanh(\cdot)$ in $[-1, 1]$ could limit the expressiveness of the model, which can be alleviated by introducing a non-linear gating mechanism. Thus, in Eq.~\eqref{eq.v.s.alpha}, a gate function $\sigma(\mat{\hat{W}}_{s}\mat{\tilde{s}}_{i}^{v})$ is incorporated, which is indeed found effective in our experiments.

\subsubsection{Sentence Aggregator for Users.}
Next, we develop an interaction-based sentence aggregator for users. Given a user--item pair, we aim to select a homogeneous subset of sentences from each of the user's reviews such that the selected sentences are relevant to the item to be recommended, i.e., the {\em target item}. In the following, we introduce a co-attentive network that uses the target item's sentences to guide the search of user's sentences.%

After the sentence encoding layer, we can represent each review by a matrix $\mat{R} = [\mat{\tilde{s}}_{1}; ...; \mat{\tilde{s}}_{k}] \in \mathbb{R}^{d \times k}$, where $[\cdot ; \cdot]$ is the concatenation operation. Suppose a user has $n$ reviews and an item has $m$ reviews. Our method first concatenates all sentences of the item to form $[\mat{R}_{1}^{v}; ...; \mat{R}_{m}^{v}] \in \mathbb{R}^{d \times mk}$, whose constituent sentences are all relevant to the target item, and thus can be used to guide the search of similar sentences from the user's reviews. To this end, we iterate over each $\mat{R}_{i}^{u}$ ($1 \le i \le n$) to calculate an affinity matrix as follows, where the superscript $u$ indicates the user notation.
\begin{equation}\label{eq.s.g}
    \begin{aligned}
        \mat{G}_{i} = \phi(f(\mat{R}_{i}^{u})^{\top}\mat{M}_{s}f([\mat{R}_{1}^{v}; ...; \mat{R}_{m}^{v}])),~~(1 \le i \le n)
    \end{aligned}
\end{equation}
Here, $\mat{M}_{s} \in \mathbb{R}^{d_{s} \times d_{s}}$ is a learnable parameter, $\phi(\cdot)$ is an activation function such as ReLU, and $f(\cdot)$ is a mapping function such as a multi-layer perceptron (MLP). If $f(\cdot)$ is the identity mapping, Eq.~\eqref{eq.s.g} becomes a bilinear mapping. Here, the $(p, q)$-th entry of $\mat{G}_{i}$ represents the affinity between the $p$-th sentence of $\mat{R}_{i}^{u}$ and the $q$-th sentence of $[\mat{R}_{1}^{v}; ...; \mat{R}_{m}^{v}]$.

To measure how relevant the $p$-th sentence of the user's review $\mat{R}_{i}^{u}$ is to the target item, we use the maximum value in the $p$-th row of $\mat{G}_{i}$. The intuition is that, if a user's sentence (i.e., a row of $\mat{G}_{i}$) has a large affinity to at least one sentence of the target item (i.e., a column of $\mat{G}_{i}$) -- in other words, the maximal affinity of this row is large -- then this user's sentence is relevant to the target item.%

However, not all sentences of the target item are useful for searching relevant sentences from the user. For instance, in Fig.~\ref{fig.intro}, the first sentence of the item's review 2, ``\emph{I received it three days ago.}'', conveys little information about the target item, and hence cannot aid in identifying relevant sentences from the user, and indeed may introduce noise into the affinity matrix. To solve this problem, recall that $\alpha_{i}^{v}$ in Eq.~\eqref{eq.v.s.alpha} represents how informative an item's sentence is. Thus, we concatenate $\alpha_{i}^{v}$'s of all sentences of the target item to form $\boldsymbol{\alpha}^{v} \in \mathbb{R}^{1 \times mk}$. Subsequently, we compute an element-wise product between each row of $\mat{G}_{i}$ and the vector $\boldsymbol{\alpha}^{v}$, i.e., $\mat{G}_{i}\otimes_{\text{row}}\boldsymbol{\alpha}^{v}$. In this manner, the $(p, q)$-th entry, $(\mat{G}_{i} \otimes_{\text{row}} \boldsymbol{\alpha}^{v})_{pq}$, is high only if the $p$-th sentence of the user is similar to the $q$-th sentence of the target item and the $q$-th sentence of the target item is non-trivial.

By summarizing the above insights, we learn attention weights for the sentences in $\mat{R}_{i}^{u}$ for each $i \in [1, n]$ by
\begin{equation}\label{eq.s.coatt}
    \begin{aligned}
        \boldsymbol{\alpha}_{i}^{u} = \text{softmax}(\text{max}_{\text{row}}(\mat{G}_{i} \otimes_{\text{row}} \boldsymbol{\alpha}^{v})),
    \end{aligned}
\end{equation}
where $\max_{\text{row}}$ refers to row-wise max-pooling for obtaining the maximum affinity. Intuitively, $(\boldsymbol{\alpha}_{i}^{u})_{j}$ is large if the $j$-th sentence in the $i$-th review of the user describes some aspects of some item that is highly similar to the target item. This serves our purpose for selecting a homogeneous subset of sentences from the user.

Next, we use $\boldsymbol{\alpha}_{i}^{u}$ to aggregate the sentences in $\mat{R}_{i}^{u}$ to infer an embedding of the $i$-th review for the user:
\begin{equation}\label{eq.r.u}
    \begin{aligned}
        \mat{r}_{i}^{u} = \sum_{j=1}^{k}(\boldsymbol{\alpha}_{i}^{u})_{j}(\mat{R}_{i}^{u})_{*j},
    \end{aligned}
\end{equation}
where $(\mat{R}_{i}^{u})_{*j}$ is the $j$-th column of $\mat{R}_{i}^{u}$. Recall that $\mat{R}_{i}^{u} = [\mat{\tilde{s}}_{1}^{u}; ...; \mat{\tilde{s}}_{k}^{u}]$, where each column of $\mat{R}_{i}^{u}$ is a sentence embedding. Note that our method iterates over $i$ for $i \in [1, n]$ to calculate all review embeddings $\mat{r}_{1}^{u}$, ..., $\mat{r}_{n}^{u}$.

\subsubsection{Remark.} Our co-attentive mechanism employs the idea of sequence pair modeling but notably differs from the conventional co-attention used in QA systems \cite{santos2016attentive,xiong2016dynamic,zhang2017attentive}. First, we only consider one side of the affinity matrix, i.e., the user. Second, our affinity matrix is adapted by row-wise multiplication of $\boldsymbol{\alpha}^{v}$ to quantify the utility of the item's sentences. Thus, our method is designed specifically for learning asymmetric attentions from user--item interactions.

\subsection{Review-Level Aggregation}

From Eq.~\eqref{eq.r.v}, we obtain review embeddings for an item, $\mat{r}_{1}^{v}$, ..., $\mat{r}_{m}^{v}$. From Eq.~\eqref{eq.r.u}, we obtain review embeddings for a user, $\mat{r}_{1}^{u}$, ..., $\mat{r}_{n}^{u}$. As shown in Fig.~\ref{fig.model}, based on these review embeddings, we develop review-level aggregators to infer an embedding for each user and item, respectively.

As discussed before, different reviews exhibit different degrees of informativeness in modeling users and items. In particular, an item's reviews are homogeneous. Thus, we are interested in reviews with rich descriptions regarding its relevant aspects and corresponding sentiments, such as the reviews 1--3 of $v$ in Fig.~\ref{fig.intro}, compared with the less informative review 4 of $v$. To attend to such reviews, similar to Eq.~\eqref{eq.r.v}, we aggregate the review embeddings to represent an item by
\begin{equation}\label{eq.v}
    \begin{aligned}
        \mat{\tilde{v}} = \sum_{i=1}^{m}\beta_{i}^{v}\mat{r}_{i}^{v},
    \end{aligned}
\end{equation}
where $\sum_{i=1}^{m}\beta_{i}^{v} = 1$, and $\beta_{i}^{v}$ is the attention weight assigned to review $\mat{r}_{i}^{v}$. It quantifies the informativeness of the review $\mat{r}_{i}^{v}$ with respect to $v$'s overall rating. $\beta_{i}^{v}$ is produced by an attentive module with gating mechanism as follows:
\begin{equation}\label{eq.v.r.beta}
    \begin{aligned}
        \beta_{i}^{v} = \frac{\exp{(\mat{v}_{r}^{\top}(\tanh{(\mat{W}_{r}\mat{r}_{i}^{v})}\otimes\sigma(\mat{\hat{W}}_{r}\mat{r}_{i}^{v})))}}{\sum_{j=1}^{k}\exp{(\mat{v}_{r}^{\top}(\tanh{(\mat{W}_{r}\mat{r}_{j}^{v})}\otimes\sigma(\mat{\hat{W}}_{r}\mat{r}_{i}^{v})))}},
    \end{aligned}
\end{equation}
where $\mat{v}_{r} \in \mathbb{R}^{h \times 1}$, $\mat{W}_{r} \in \mathbb{R}^{h \times d}$, and $\mat{\hat{W}}_{r} \in \mathbb{R}^{h \times d}$ are model parameters.

At the same time, a user's reviews are heterogeneous concerning a variety of items that the user has purchased, and not all reviews are relevant to the target item. Thus, similar to Eq.~\eqref{eq.s.g} and Eq.~\eqref{eq.s.coatt}, given a user--item pair, a review-level co-attentive network is designed to select reviews from the user as guided by the reviews of the item.

Specifically, an affinity matrix at the review level 
\begin{equation}\label{eq.r.g}
    \begin{aligned}
        \mat{G} = \phi(f([\mat{r}_{1}^{u}; ...; \mat{r}_{n}^{u}])^{\top}\mat{M}_{r}f([\mat{r}_{1}^{v}; ...; \mat{r}_{m}^{v}])),
    \end{aligned}
\end{equation}
is computed, where $\mat{M}_{r} \in \mathbb{R}^{d_{r} \times d_{r}}$ is a learnable parameter. Here, the $(p, q)$-th entry of $\mat{G}$ represents the affinity between the $p$-th review of the user and the $q$-th review of the item.

Then, attention weights for the reviews of the user 
\begin{equation}\label{eq.r.coatt}
    \begin{aligned}
        \boldsymbol{\beta}^{u} = \text{softmax}(\text{max}_{\text{row}}(\mat{G} \otimes_{\text{row}} \boldsymbol{\beta}^{v})),
    \end{aligned}
\end{equation}
are obtained, where $\boldsymbol{\beta}^{v} = [\beta_{1}^{v}, ..., \beta_{m}^{v}]$ was obtained by Eq.~\eqref{eq.v.r.beta} for the item. It is introduced to adapt $\mat{G}$ to encode important reviews of the item. Finally, we aggregate the review embeddings to represent a user by the following weighted sum.
\begin{equation}\label{eq.u}
    \begin{aligned}
        \mat{\tilde{u}} = \sum_{i=1}^{n}\beta_{i}^{u}\mat{r}_{i}^{u}
    \end{aligned}
\end{equation}

\subsubsection{Encoding Latent Rating Patterns.}
Although the embeddings $\mat{\tilde{u}}$ and $\mat{\tilde{v}}$ contain rich semantic information from reviews, there are some latent characteristics of users (items) that are not encoded by their reviews, but can be inferred from the rating patterns. For instance, a picky user might tend to uniformly pick lower ratings than a more easygoing user. To encode such personalized preferences, as inspired by \citet{koren2009matrix}, we embed a one-hot representation of the ID of each user (item) using an MLP, and obtain an embedding vector $\mat{\hat{u}}$ ($\mat{\hat{v}}$) for the user (item). This vector directly correlates with the ratings of a user (item), and is thus able to capture the latent rating patterns. Then, as illustrated in Fig.~\ref{fig.model}, we concatenate $\mat{\tilde{u}}$ and $\mat{\hat{u}}$ to obtain the final embedding of a user, i.e., $\mat{u} = [\mat{\tilde{u}}; \mat{\hat{u}}]$, and concatenate $\mat{\tilde{v}}$ and $\mat{\hat{v}}$ to obtain the final embedding of an item, i.e., $\mat{v} = [\mat{\tilde{v}}; \mat{\hat{v}}]$.

\subsection{Prediction Layer}

As shown by the top part of Fig.~\ref{fig.model}, the prediction layer receives $\mat{u}$ and $\mat{v}$, and concatenates them to $[\mat{u}; \mat{v}]$, which is then fed into a function $g(\cdot)$ to predict the rating. In this work, we realize $g(\cdot)$ as a parameterized factorization machine (FM) \cite{rendle2010factorization}, which is effective to model the pairwise interactions between the input features for improving recommendation performance. Given an input $\mat{x} \in \mathbb{R}^{d \times 1}$, $g(\cdot)$ is defined as
\begin{equation}
    \begin{aligned}
        g(\mat{x}) = b + \sum_{i=1}^{d}\mat{w}_{i}\mat{x}_{i} + \sum_{i=1}^{d}\sum_{j=i+1}^{d}\braket{\mat{z}_{i}, \mat{z}_{j}}\mat{x}_{i}\mat{x}_{j},
    \end{aligned}
\end{equation}
where $b$ is a bias term, $\mat{w}$ is a parameter for linear regression, $\{\mat{z}_{i}\}_{i=1}^{d}$ are the factorized parameter for modeling the pairwise interactions between $\mat{x}_{i}$ and $\mat{x}_{j}$, $\braket{\cdot, \cdot}$ denotes the inner product, and the output of $g(\mat{x})$ is the predicted rating.

To learn model parameters, we minimize the difference between the true ratings and the predicted ratings, as measured by the mean squared error
\begin{equation}\label{eq.l}
    \begin{aligned}
        \ell = \frac{1}{c}\sum_{i=1}^{c}(y_{i} - g([\mat{u}; \mat{v}]))^{2},
    \end{aligned}
\end{equation}
where $c$ is the total number of user--item pairs in the training data, and $y_{i}$ is the true rating of the $i$-th user--item pair. The $\ell$ in Eq.~\eqref{eq.l} serves as our loss function for model training.

\section{Experiments}

In this section, we evaluate our \ourmethod\ model on several real datasets and compare it with state-of-the-art approaches.

\subsection{Datasets}

\begin{table}[!t]
\caption{Statistics of datasets}\label{tab.data}
\footnotesize
\begin{center}
\begin{tabular}{|l|l|l|l|} \hline
{\bf Dataset} & {\bf \#Users} & {\bf \#Items} & {\bf \#Reviews} \\ \hline
Digital\_Music (DM) & 5,541& 3,568& 64,706\\ \hline
Office\_Products (OP) &4,905 & 2,420& 53,258\\ \hline
Health (HE) &38,609 &18,534 &346,355 \\ \hline
Toys\_and\_Games (TG) & 19,412&11,924 & 167,597\\ \hline
Kindle\_Store (KS) & 68,223&61,935 & 982,619\\ \hline
Pets\_Supplies (PS) & 19,856&8,510 & 157,836\\ \hline
Tools\_and\_Home (TH) & 16,638 & 10,217 & 134,476\\ \hline
Videos\_Games (VG) & 24,303&10,672 & 231,780 \\ \hline
Automotive (AM) & 2,928& 1,835&20,473 \\ \hline
Yelp &88,370 & 33,902 & 1,332,447\\ \hline
\end{tabular}
\end{center}
\end{table}

\begin{table*}[!t]
\caption{MSE results of the compared methods on different 5-core datasets}\label{tab.result.5core}
\footnotesize
\begin{center}
\begin{tabular}{|l|c|c|c|c|c|c|c|c|c|} \hline
Dataset & FM & PMF & NMF & SVD & DeepCoNN & D-ATT & MPCN & HUITA & \ourmethod\\ \hline
Digital\_Music (DM) & 0.8498&0.8788&1.0491 & 1.0843 & 0.8754& 0.8506&\underline{0.8396}&0.8719 &\bf{0.8172}\\ \hline
Office\_Products (OP) &0.7291 &0.7807& 0.9285&0.7906& 0.7253& 0.7124& 0.7084& \underline{0.7082}& \bf{0.6825}\\ \hline
Health (HE) &1.1825 & 1.2076&1.4317 &1.1508 & 1.0862& 1.0915& \underline{1.0817}&1.1207&\bf{1.0743}\\ \hline
Toys\_and\_Games (TG) & 0.8639&0.9192 &1.1105 &0.9188 & 0.8391&\underline{ 0.8364}&0.8452& 0.8969& \bf{0.8220}\\ \hline
Kindle\_Store (KS) &0.6469  &0.6695 &0.8032  & 0.7370 &0.6514 &\underline{0.6382} &0.6577 &0.6544 & \bf{0.6270}\\ \hline
Pets\_Supplies (PS) &1.3303  &1.434 & 1.6806 & 1.3191 &1.2598 &1.2730 &\underline{ 1.2566} &1.3038 & \bf{1.2515}\\ \hline
Tools\_and\_Home (TH)&1.0229  & 1.1182&1.3580  &1.0373 &0.9856 &\underline{0.9850} &0.9871  &1.0189 & \bf{0.9671}\\ \hline
Videos\_Games (VG) &1.1849  &1.2473 &1.4357  & 1.3168 &1.1575 &\underline{1.1448} &1.1747 & 1.1772& \bf{1.1138}\\ \hline
Automotive (AM) & 0.8189&0.9187 &1.2074 & 0.8140 &0.7809 & 0.7654&\underline{0.7643}&0.7766 &\bf{0.7314}\\ \hline
Yelp &1.6094 &1.8207 & 1.8389 & 1.6615  & \underline{1.5957}  &1.5959  & 1.6195  & 1.6105 &\bf{1.5735}\\ \hline
\end{tabular}
\end{center}
\end{table*}

\begin{table*}[!t]
\caption{MSE results of the compared methods on different 10-core datasets}\label{tab.result.10core}
\footnotesize
\begin{center}
\begin{tabular}{|l|c|c|c|c|c|c|c|c|c|} \hline
Dataset & FM & PMF & NMF & SVD & DeepCoNN & D-ATT & MPCN & HUITA & \ourmethod\\ \hline
Digital\_Music (DM) & 0.8611 & 0.8641  &  0.9491 &  0.8503& 0.8734& \underline{0.8429}  & 0.8629 & 0.8512 &\bf{ 0.7880}  \\ \hline
Office\_Products (OP) & 0.6291 &0.6695 &0.7346  &0.6757 &0.6016 &\underline{0.5914}  &0.6120  & 0.6009 & \bf{0.5717}\\ \hline
Health (HE) & 0.8166 & 0.9158 & 0.9200 &0.8275  & 0.8328 & \underline{0.8019} &0.8020 &0.8177& \bf{0.7802}  \\ \hline
Toy\_and\_Games (TG) &0.6904  & 0.6233 & 0.7575 & 0.6331 &0.6331 & \underline{0.6292} & 0.6412 &0.6303 & \bf{0.5964}  \\ \hline
Kindle\_Store (KS) & 0.5954 &0.6035 & 0.6305 & 0.6483 & 0.5325& 0.5275 &\underline{ 0.5124} & 0.5312& \bf{0.5092}\\ \hline
Pet\_Supplies (PS) &1.2236  &1.5239  &1.2536  & 0.9950 &0.9927 & \underline{0.9616} & 1.0722 &1.0168 & \bf{0.9421}  \\ \hline
Tools\_and\_Home (TH)& 0.8746 &0.7668  & 0.9032  & 0.7391 &0.6632 & \underline{0.6297}  & 0.6507 &0.6445 & \bf{0.5948}\\ \hline
Videos\_Games (VG) &1.0611 &1.0718 &1.2435  &1.0318   & 1.0743 & \underline{1.0365} & 1.0730 &1.0697  &\bf{0.9927} \\ \hline
Yelp &1.5432 & 1.4734 &1.5735  & 1.4025 & \underline{1.3961}  & 1.4018 &1.4033 &1.4040  &\bf{1.3671}\\ \hline
\end{tabular}
\end{center}
\end{table*}

\begin{figure*}[!t]
\centerline
{
\subfigure{\includegraphics[width=0.47\columnwidth]{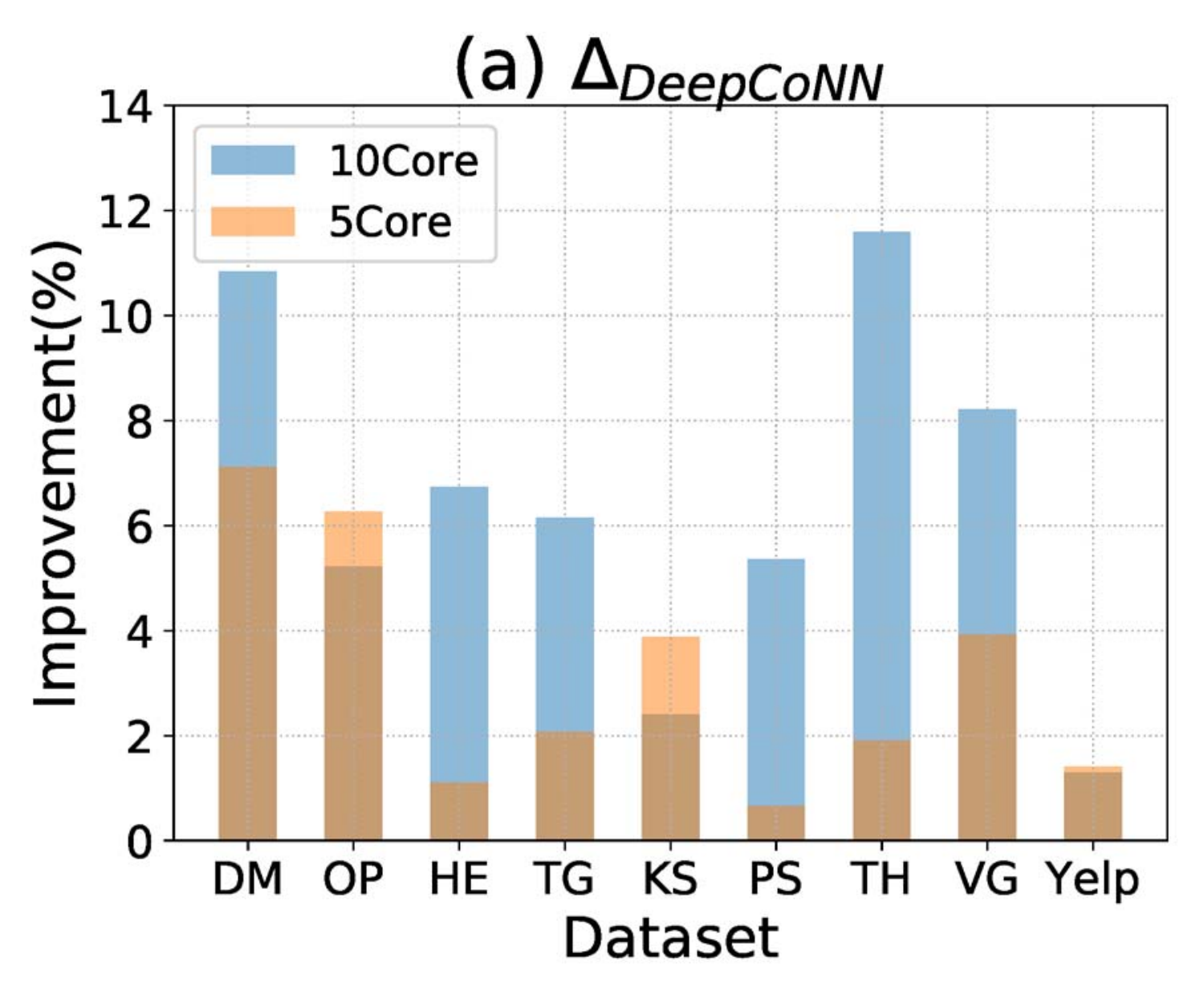}}
\subfigure{\includegraphics[width=0.47\columnwidth]{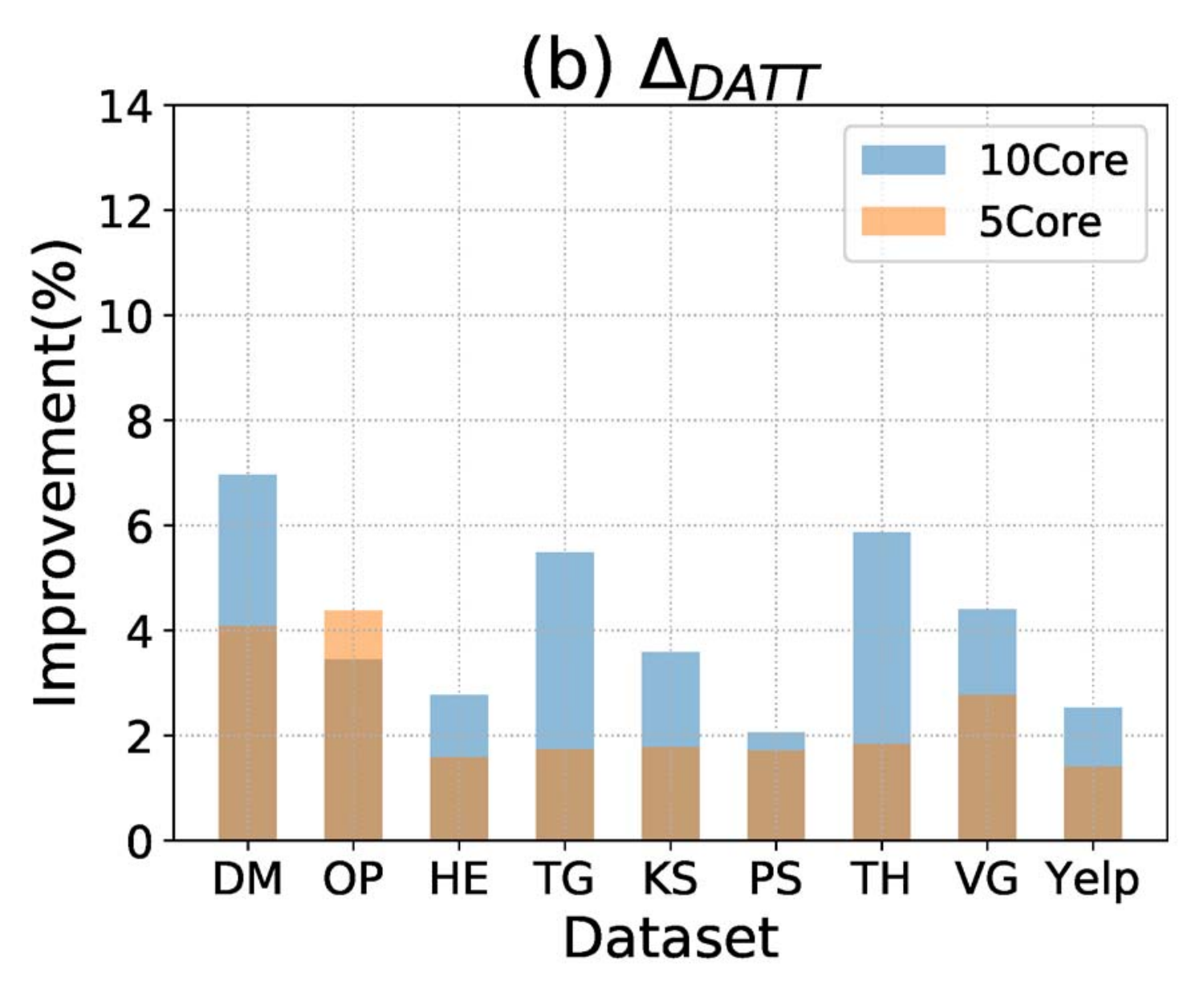}}
\subfigure{\includegraphics[width=0.47\columnwidth]{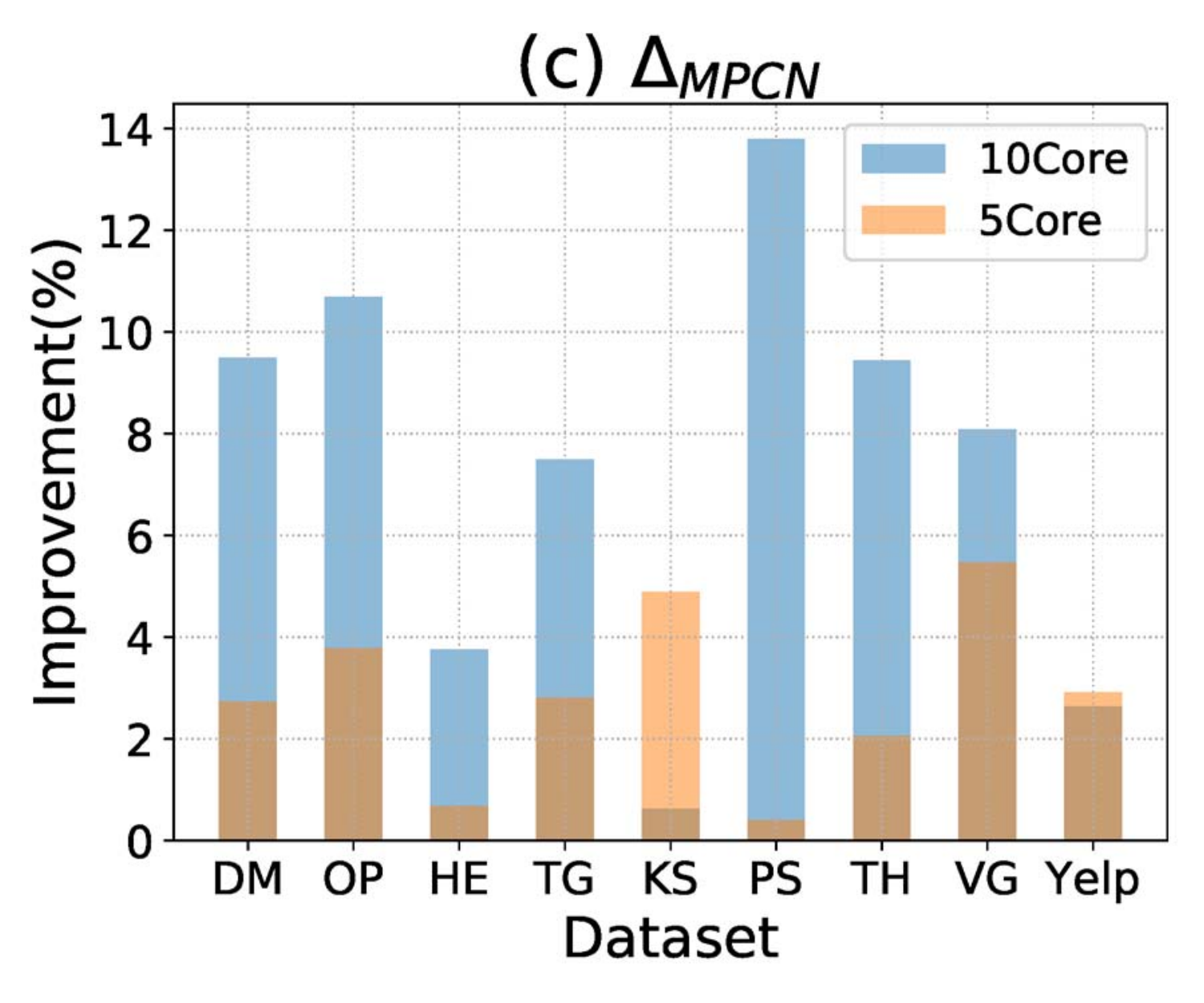}}
\subfigure{\includegraphics[width=0.47\columnwidth]{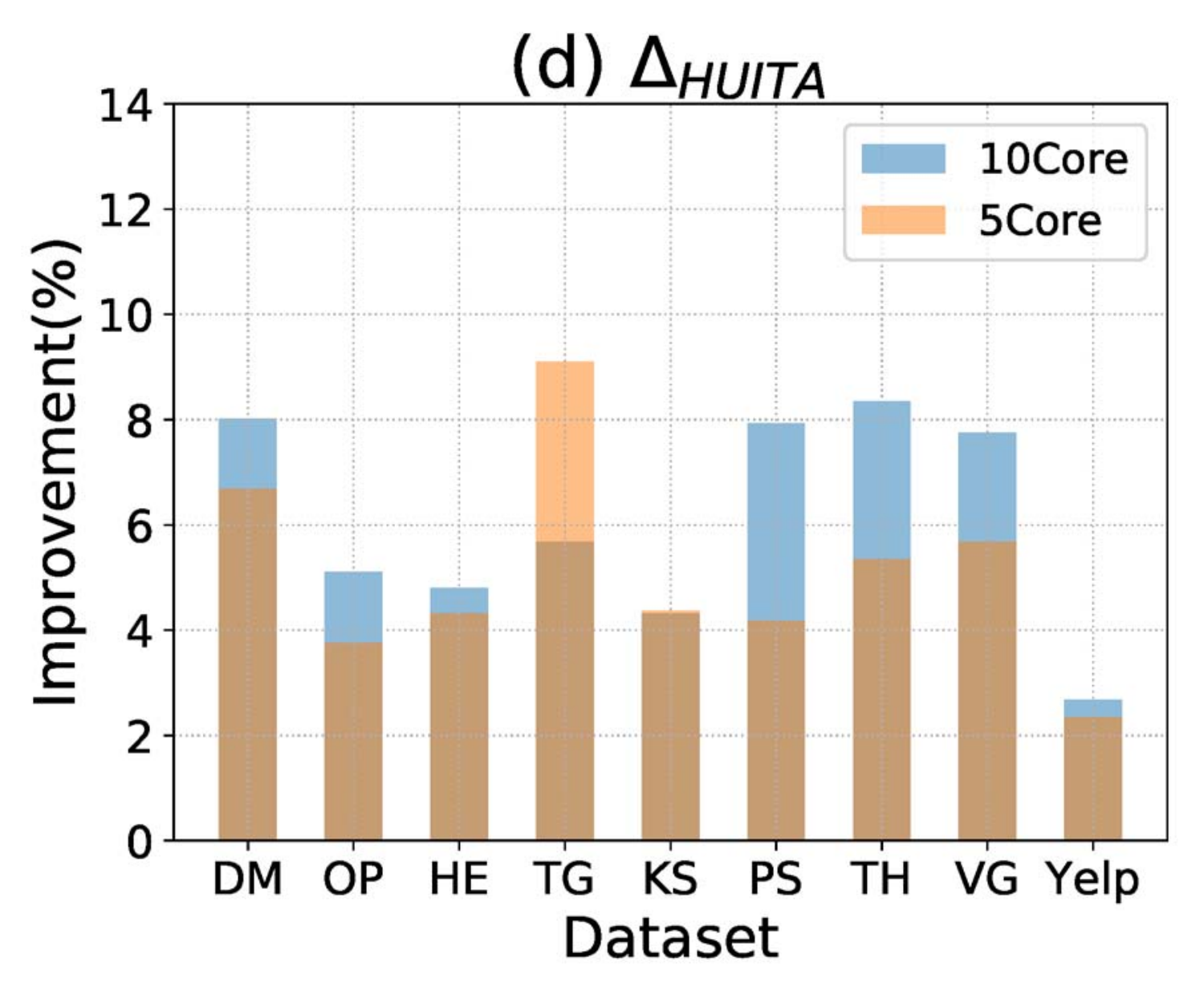}}
}

\caption{The relative improvements of \ourmethod\ over (a) DeepCoNN, (b) D-ATT, (c) MPCN, and (d) HUITA, on different datasets. The abbreviations of the datasets can be found in Table \ref{tab.data}-\ref{tab.result.10core}. Here, blue refers to the improvement on the 10-core datasets, orange refers to the improvement on the 5-core datasets, brown is the overlapped area between blue and orange.}\label{fig.improvement}
\end{figure*}

We conducted experiments on $10$ different datasets, including $9$ Amazon product review datasets for $9$ different domains, and the large-scale Yelp challenge dataset\footnote{https://www.yelp.com/dataset/challenge} on restaurant reviews. Table \ref{tab.data} summarizes the domains and statistics for these datasets. Across all datasets, we follow the existing work \cite{seo2017interpretable,tay2018multi} to perform preprocessing to ensure they are in a $t$-core fashion, i.e., the datasets only include users and items that have at least $t$ reviews. In our experiments, we evaluate the two cases of $t=5$ and $t=10$. For the Yelp dataset, we follow \citet{seo2017interpretable} to focus on restaurants in the AZ metropolitan area. For each dataset, we randomly split the user--item pairs into $80\%$ training set, $10\%$ validation set, and $10\%$ testing set. When learning the representations for users and items, we only use their reviews from the training set, and none from the validation and testing sets. This ensures a practical scenario where we cannot include any future reviews into a user's (item's) history for model training. %

\subsection{Compared Methods}

We compare our model with both conventional approaches and state-of-the-art approaches, including Factorization Machines (FM) \cite{rendle2010factorization}, SVD \cite{koren2009matrix}, Probabilistic Matrix Factorization (PMF) \cite{mnih2008probabilistic}, Nonnegative Matrix Factorization (NMF) \cite{lee2001algorithms}, DeepCoNN \cite{zheng2017joint}, D-ATT \cite{seo2017interpretable}, MPCN \cite{tay2018multi}, and HUITA \cite{wu2019hierarchical}.

Among these methods, FM, SVD, PMF, and NMF are rating-based collaborative filtering methods. DeepCoNN, D-ATT, MPCN, and HUITA are state-of-the-art methods that leverage the semantic information in reviews for improved performance. Specifically, DeepCoNN uses the same CNN module to learn user and item embeddings based on their reviews for recommendation. D-ATT extends DeepCoNN by adding a dual-attention layer at word-level before convolution. MPCN %
attends to informative reviews by several pointers.
HUITA uses a symmetric hierarchical structure to infer user (item) embeddings using regular attention mechanisms. It is worth noting that all of the above review-based methods regard user reviews and item reviews as the same type of documents and process them in an identical way. 

Finally, to gain further insights on some of the design choices of our \ourmethod\ model, we compare \ourmethod\ with its variants, which will be discussed later in the ablation analysis.

\subsection{Experimental Settings}

The parameters of the compared methods are selected based on their performance on the validation set. Specifically, for FM, the dimensionality of the factorized parameters is 10. For SVD, PMF, and NMF, the number of factors is set to 50. DeepCoNN uses 100 convolutional kernels with window size 3. D-ATT uses 200 filters and window size 5 for local attention; 100 filters and window sizes [2, 3, 4] for global attention. MPCN uses 3 pointers, and hidden dimensionality of 300 for inferring affinity matrix. HUITA uses 200 filters in the word-level CNN with window size 3, and 100 filters in the sentence-level CNN with window size 3.

For our \ourmethod\ model, the dimensionality of the hidden states of the BiLSTM is set to 150. The dimensionality of the user and item ID embeddings are set to 300. The dimensionality of $\mat{M}_{s}$ ($\mat{M}_{r}$) in Eq.~\eqref{eq.s.g} (Eq.~\eqref{eq.r.g}) is 300. We apply dropout \cite{srivastava2014dropout} with rate $0.5$ after the fully connected layer to alleviate the overfitting problem. The loss function is optimized by Adam \cite{kingma2014adam}, with a learning rate of 0.0002 and a maximum of 10 epochs.

For the methods DeepCoNN, D-ATT, and HUITA, the pre-trained GloVe \cite{pennington2014glove} are used to initialize the word embeddings. For MPCN and our \ourmethod, the word embeddings are learned from scratch since using pre-trained embeddings generally degrades their performance. For all methods, the dimensionality of the word embedding is set to 300. %
We independently repeat each experiment $5$ times, and use the averaged mean square error (MSE) \cite{zheng2017joint} to quantitatively evaluate the performance.

\subsection{Experimental Results}

Table \ref{tab.result.5core} summarizes the results of the compared approaches on the 5-core datasets. We have several observations from the results. First, review-based methods generally outperform rating-based methods. %
This validates the usefulness of reviews in providing fine-grained information for refining user and item embeddings for improving the accuracy of rating prediction. Second, methods that distinguish reviews, such as D-ATT and MPCN, often outperform DeepCoNN, which suggests that different reviews exhibit different degrees of importance for modeling users and items. We also observe that HUITA does not show superiority over DeepCoNN. This may stem from its symmetric style of attention learning, which does not make much sense when reviews are heterogeneous. Finally, the proposed \ourmethod\ consistently outperforms other methods, which demonstrates the effectiveness of distinguishing the learning of user and item embeddings via asymmetric attentive modules so as to infer more reasonable attention weights for recommendation.

Table \ref{tab.result.10core} presents the results on the 10-core datasets, from which the {\em Automotive} dataset is excluded because only very few users and items are left after applying the 10-core criterion on it. In contrast to Table \ref{tab.result.5core}, %
all methods in general achieve better results in Table \ref{tab.result.10core}, since more ratings and reviews become available for each user and item. In this case, we observe that D-ATT often outperforms MPCN. This may be because the Gumbel-Softmax pointers in MPCN make hard selections on reviews, thereby filtering out many reviews that may result in a significant loss of information. This problem is more severe when users (items) have more useful reviews, as in the 10-core scenario. Additionally, we observe that the performance gaps between \ourmethod\ and the compared methods become larger. Specifically, summarizing the relative improvements of \ourmethod\ over each of the review-based methods in Fig.~\ref{fig.improvement}, 
we observe that \ourmethod\ generally gains more on the 10-core datasets, with absolute gains of up to $11.6\%$ (DeepCoNN), $7.0\%$ (D-ATT), $13.8\%$ (MPCN), and $8.4\%$ (HUITA). This suggests that the more reviews each user and item has, the more important it is to perform proper attention learning on relevant reviews and sentences on both the user and item sides.

\subsection{Case Study}

\begin{figure}[!t]
\includegraphics[width=1.0\columnwidth]{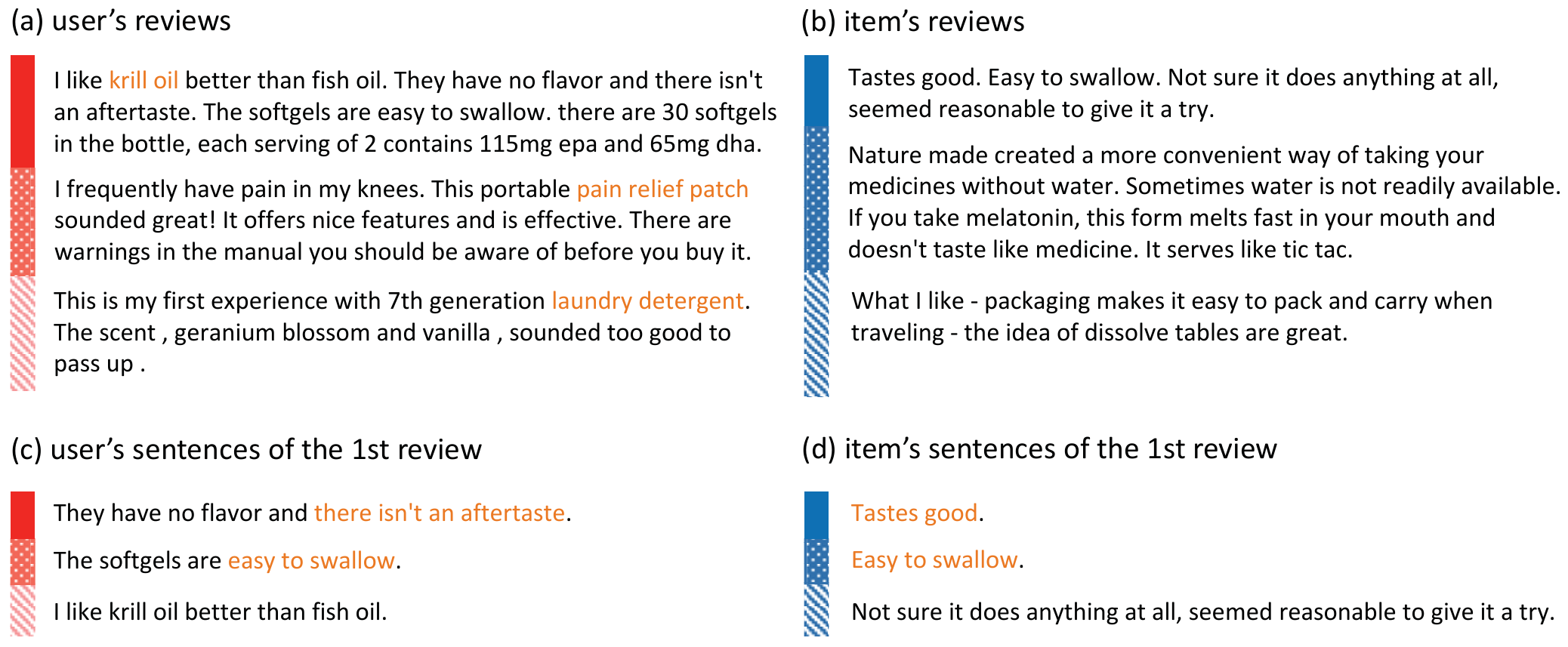}
\caption{The visualization of attention weights on (a) user's reviews, (b) item's reviews, (c) user's sentences, (d) item's sentences. The item is a sleep aid medicine. The vertical bars represent weights. Darker colors indicate higher weights.}\label{fig.casestudy}
\end{figure}

We next investigate the interpretability of \ourmethod. Fig.~\ref{fig.casestudy}(a) and (b) show the attention weights of \ourmethod\ on the top three reviews of a pair of user and item on the {\em Health} dataset, where the item is a sleep aid medicine. In each of the user's reviews, the highlighted words indicate the item described by the review. As can be seen, the first two items ``krill oil'' and ``pain relief patch'' are more relevant to the item ``sleep aid medicine'' than the ``laundry detergent'' in the lowest-weighted review. On the other hand, the top two reviews of the item are more informative with regard to the aspects of the item than the last review, which only discusses packaging, a rather marginal aspect of medication. Thus, the review-level attention weights of \ourmethod\ are meaningful.

Fig.~\ref{fig.casestudy}(c) and (d) zoom into the attention weights of \ourmethod\ on the top three sentences of the first review of the user and item, respectively. The highlighted words indicate the reason of why the sentences are ranked highly. Apparently, the user cares about the taste of the medicine and prefers easily-swallowed softgels, while the item indeed appears to taste good and is easy to swallow. Although the first two sentences in Fig.~\ref{fig.casestudy}(d) are short, they convey more useful information than the lowest-weighted sentence. Thus, the sentence-level attention weights 
are also meaningful. This explains why \ourmethod\ predicts a 4.4 rating score on this user--item pair, close to the true rating 5.0 given by the user.

\subsection{Ablation Analysis}
\label{subsec:ablation}

\begin{table}[!t]
\caption{Ablation analysis}\label{tab.abl}
\footnotesize
\begin{center}
\begin{tabular}{|l|c|c|c|c|} \hline
Model & VG & DM & AM  & OP\\ \hline
\ourmethod & \bf{1.1138} & \bf{0.8172} & \bf{0.7314} & \bf{0.6825}\\
(a) --Item aggregators &1.1286 &0.8205 &0.7506 &0.6951\\
(b) --User aggregators &1.1604 &0.8246 &0.7467 &0.6941\\
(c) --Adapted affinity &1.1363 &0.8229&0.7348 &0.6936\\
(d) --FM &1.1267 &0.8341 &0.7723 &0.7078\\
(e) --Gating &1.1220 &0.8188 &0.7385 &0.6883\\ \hline
\end{tabular}
\end{center}
\end{table}

Table \ref{tab.abl} presents the results of our ablation analysis using four datasets. In the table, \ourmethod\ is our original model. In (a), the item's attention modules are replaced by average-pooling. In (b), the user co-attention modules are replaced by attention modules similar to the item ones and this thus constitutes a symmetric model. In (c), we remove the row-wise multiplication between the affinity matrix and the attention weights in Eqs.~\eqref{eq.s.coatt} and \eqref{eq.r.coatt}. In (d), the parameterized factorization machine is replaced by a dot product. In (e), the gating mechanisms in Eqs.~\eqref{eq.v.s.alpha} and \eqref{eq.v.r.beta} are removed.

From Table \ref{tab.abl}, we observe that different variants of \ourmethod\ show suboptimal results to various degrees. Comparing with (a), we can observe the importance of considering attention weights on the sentences and reviews of each item. The degraded MSEs of (b) suggest that our asymmetric design in the model architecture is essential. The results of (c) validate our design of the attention-adapted affinity matrix in Eqs.~\eqref{eq.s.coatt} and \eqref{eq.r.coatt}. The substantial MSE drops for (d) establish the superiority of using FM as the prediction layer. The comparison between (e) and \ourmethod\ suggests the effectiveness of the gating mechanisms. Thus, the results of the ablation study validate the design choices of our model architecture.

\section{Conclusions}
In this work, we highlight the asymmetric attention problem for review-based recommendation, which has been ignored by existing approaches. To address it, we propose a flexible neural architecture, \ourmethod, which is characterized by its asymmetric attentive modules for distinguishing the learning of user embeddings and item embeddings from reviews, as well as by its hierarchical paradigm to extract fine-grained signals from sentences and reviews. Extensive experimental results on datasets from different domains demonstrate the effectiveness and interpretability of our method.

\bibliographystyle{aaai}
\bibliography{ref}

\end{document}